\documentclass[a4paper,cits]{PoS}
\usepackage{amsmath,amssymb,amsfonts} 

\usepackage[normalem]{ulem}
\usepackage[utf8]{inputenc}
\usepackage[T1]{fontenc}

\usepackage{cancel}
\usepackage{graphicx}
\usepackage{latexsym}
\usepackage{amsmath}
\usepackage{amssymb}
\usepackage{bbm}
\usepackage{pdfsync}
\usepackage{epsfig}
\usepackage{epstopdf}
\usepackage{subfigure}
\usepackage{color}
\usepackage{comment}

\usepackage{slashed}
\usepackage{placeins}
\usepackage{cleveref}
\usepackage{setspace}

\newcommand{\gGF}{g_{\rm GF}}
\newcommand{\MSb}{\overline{\textrm{MS}}}

\title{Infrared Behaviour of SU(2) Gauge Theory with $N_f$ fundamental flavours}

\ShortTitle{Infrared Behaviour of SU(2) Gauge Theory}

\author{\speaker{Viljami Leino}\\ Technische Universit\"at M\"unchen, Physik Department T30f, James-Franck-Str. 1, 85748
Garching, Germany \\ E-mail: \email{viljami.leino@tum.de}}

\author{Kari Rummukainen\\
		Helsinki Institute of Physics and Department
		of Physics, University of Helsinki\\ E-mail:
		\email{kari.rummukainen@helsinki.fi}}

\author{Joni M. Suorsa\\
		Helsinki Institute of Physics and Department
		of Physics, University of Helsinki} %\\ E-mail: \email{joni.suorsa@helsinki.fi}}

\author{Kimmo Tuominen \\
		Helsinki Institute of Physics and Department
		of Physics, University of Helsinki\\ E-mail:
		\email{kimmo.i.tuominen@helsinki.fi}}

\author{Sara T\"ahtinen \\
		Helsinki Institute of Physics and Department
		of Physics, University of Helsinki\\ E-mail:
		\email{sara.tahtinen@helsinki.fi}}

\abstract{
We review our recent results on the infrared behaviour of the SU(2) gauge theory with $N_f$
massless fundamental flavour fermions. We have analyzed the running of the coupling in SU(2) gauge theories with
six and eight fermionic flavours using the gradient flow step scaling method. 
From the running of the coupling, we see a clear indication of an infrared stable fixed point
in theories with six and eight flavours. These results are confirmed by our mass spectrum study,
where we varied the number of flavours from two to six. We also compute the anomalous dimensions
of mass and coupling.
}

\FullConference{XIII Quark Confinement and the Hadron Spectrum - Confinement2018\\
		31 July - 6 August 2018\\
		Maynooth University, Ireland}

\begin{document}

\section{Introduction} 
The long distance behaviour of a strongly interacting $SU(N_c)$ gauge theory 
has a strong dependence on the number of fermion flavours $N_f$ and on the representation under which the fermions transform.
When the number of fermions is reasonably small, these theories break the chiral symmetry similarly to QCD and are confining.
However, if the number of flavours is increased while the number of colours is kept the same, the theory will develop
a non-trivial infrared fixed point (IRFP) and the long distance behaviour of the theory becomes conformal.
Increasing the number of flavours even further will cause the asymptotic freedom to be lost. This upper limit in the number of flavours
before losing the asymptotic freedom is perturbatively known to be $N_f^\mathrm{AF}=11N_c/2$ and is called the upper edge of the conformal window.
The exact number of fermions at which the conformality onsets is called the lower edge of the conformal window and is unknown. 
Different perturbative estimates have been given for the location of the lower edge, but since the conformality generally
onsets at large coupling, a non-perturbative analysis is required.

In this paper we focus on the $N_c=2$ gauge theory with 2--8 
fundamental representation Dirac fermions in the limit of vanishing quark mass 
and try to paint a complete picture of the conformal window in $SU(2)$ theories. 
These $SU(2)$ symmetric models offer a simple framework for testing different long distance dynamics~\cite{Sannino:2004qp}.
The current status of lattice simulation~\cite{Ohki:2010sr,Bursa:2010xn,Karavirta:2011zg,Hayakawa:2013maa,Appelquist:2013pqa,Leino:2017zyg,Leino:2017lpc,Leino:2017hgm,Leino:2018qvq,Amato:2018nvj}
indicate that the $N_f=2,4$ are confining and $N_f=6-10$ are inside the conformal window. Above $N_f=11$ the asymptotic freedom is lost.
Here we review our results, published in~\cite{Leino:2017lpc,Leino:2017hgm,Leino:2018qvq,Amato:2018nvj}, that clarified the existence of an IRFP in
$N_f=6$ and $N_f=8$ models.

We have performed running of the coupling analyses on $N_f=6$ and $N_f=8$ models and confirmed that they feature an IRFP. 
From the same configurations we also measure anomalous dimensions characterizing the behaviour near the IRFP.
Alongside, we have run a systematic study on the spectrum in SU(2) gauge theory with $N_f=2,4,6$.
From the scaling of hadron masses with respect to quark masses we observe chiral symmetry breaking for two and four fermion models and
a strong indication of conformality for the $N_f=6$ model. In the $N_f=6$ model we can also extract the mass anomalous dimension
which confirms the results obtained from the measurement of the running coupling.

\section{Lattice formulation}
The study of SU(2) gauge theory with $N_f$ massless Dirac fermions in the fundamental representation is carried out 
using a combination of HEX~smeared~\cite{Capitani:2006ni} $S_\mathrm{G}(V)$ and unsmeared $S_\mathrm{G}(U)$ Wilson gauge actions. 
These gauge actions are combined with a factor $c_g=0.5$.
For the Fermion action we employ clover improved Wilson fermion action $S_\mathrm{F}(V)$, with tree-level Sheikholeslami-Wohlert coefficient $c_{sw}=1$.
We use the smeared gauge links in the fermion action.
Combining these actions we get a total lattice action: $S_\mathrm{G}=(1-c_g)S_\mathrm{G}(U)+c_gS_\mathrm{G}(V)+S_\mathrm{F}(V)+c_{sw}\delta S_\mathrm{SW}(V)$.

For the running of the coupling study, we use the Schr\"odinger functional boundary conditions by having Dirichlet boundaries on temporal directions and periodic boundaries on spatial directions.
This allows us to easily tune the fermion masses to zero and subsequently measure the mass anomalous dimension. 
Elseways, for the spectrum study we choose periodic boundaries in all directions, which in turn means that the masses have to be tuned with the traditional PCAC mass relation.

In general, the simulations are run with multiple different lattice sizes and bare couplings. For exact numbers of these quantities and
comprehensive algorithmic details we refer the reader to the original papers~\cite{Leino:2017lpc,Leino:2017hgm,Leino:2018qvq,Amato:2018nvj}.

\section{Running of the coupling}\label{sec:run}
We measure the running of the coupling using the Yang-Mills gradient flow~\cite{Luscher:2009eq,Luscher:2010iy}.
With the gradient flow method we continuously smear the gauge field towards the minima of the Yang-Mills action removing the UV divergences.
The gradient flow is defined as a differential equation with respect to fictitious time $t$:
$\partial_t B_{\mu} = D_{\nu} G_{\nu\mu}$ with initial condition $B_{\mu}(x;t=0) = A_\mu(x)$. Here the 
$D_\mu=\partial_\mu+[B_\mu,\,\cdot\,]$ is the covariant derivative and $G_{\mu\nu}(x;t)$ is the field strength tensor.
On the lattice one has to choose the discretization for the $G_{\mu\nu}(x;t)$. 
We evolve the gradient flow with both with L\"uscher-Weisz action (LW) and with Wilson action (W) in order to investigate the discretization effects. 

The coupling can be measured from the evolved fields at a scale $\mu^{-1}=\sqrt{8t}$: as $\gGF^2(\mu)=\mathcal{N}^{-1}t^2\langle E(t+\tau_0)\rangle\vert_{x_0=L/2\,,\,t^{-1}=8\mu^2}$.
Here $E(t)$ is the energy density and the normalization factor $\mathcal{N}$ is defined for the Schr\"odinger functional boundary conditions in~\cite{Fritzsch:2013je}. 
%so that the gradient flow coupling agrees with $\MSb$ coupling at tree level. 
Because of the boundary conditions break the time translation invariance, we only measure the coupling along the central time slice $x_0=L/2$. 
$\tau_0$ defines an empirical $\mathcal{O}(a^2)$ improvement for the gradient flow coupling~\cite{Cheng:2014jba}.
We choose $\tau_0=0.025\mathrm{log}(1+2\gGF^2)$ for $N_f=6$~\cite{Leino:2017hgm} and $\tau_0=0.06\mathrm{log}(1+\gGF^2)$ for $N_f=8$~\cite{Leino:2017lpc}.
In~\cite{Leino:2018qvq} we employ alternative $\mathcal{O}(a^2)$ improvement by combining clover and plaquette measurements of $E(t)$ with empirical coefficient $X$~\cite{Fodor:2015baa}.
Furthermore, the fictitious time $t$ at which the coupling is measured, is determined by relating the lattice and renormalization scales with a dimensionless parameter $c_tL=8t$.
The choice of $c_t$ defines the renormalization scheme.

We measure the gradient flow coupling at multiple bare couplings $g_0^2=4/\beta$. The bare couplings are varied from 0.5 to 8 for $N_f=8$ and to 10 for $N_f=6$.
We then interpolate these couplings with either polynomial or rational ansatz. For $N_f=6$ model we perform the interpolation with 9th degree polynomial for lattices larger than $L=16$ 
and 10th degree polynomial for smaller lattices. In $N_f=8$ model we use a rational ansatz with 7th degree polynomial in the numerator and 1st degree polynomial in the denominator.
With the interpolated couplings we can define a continuous 
the step scaling function and extrapolate continuum with:
\begin{equation}\label{eq:stepscale}
\Sigma(u,L/a,s)=\gGF^2(g_0^2,sL/a)\vert_{\gGF^2(g_0^2,L/a)=u}\,,\qquad\qquad \Sigma(u,L/a)=\sigma(u)+c(u)\left(\frac{a}{l}\right)^2\,,
\end{equation}
where we have assumed discretization effects to be of order $\mathcal{O}(a^2)$.
The step scaling function allows us to evaluate the running of the coupling and the IRFP will be identified by a condition $\sigma(\gGF^2)/\gGF^2=1$.

\begin{figure}[t]
  \includegraphics[width=0.49\textwidth]{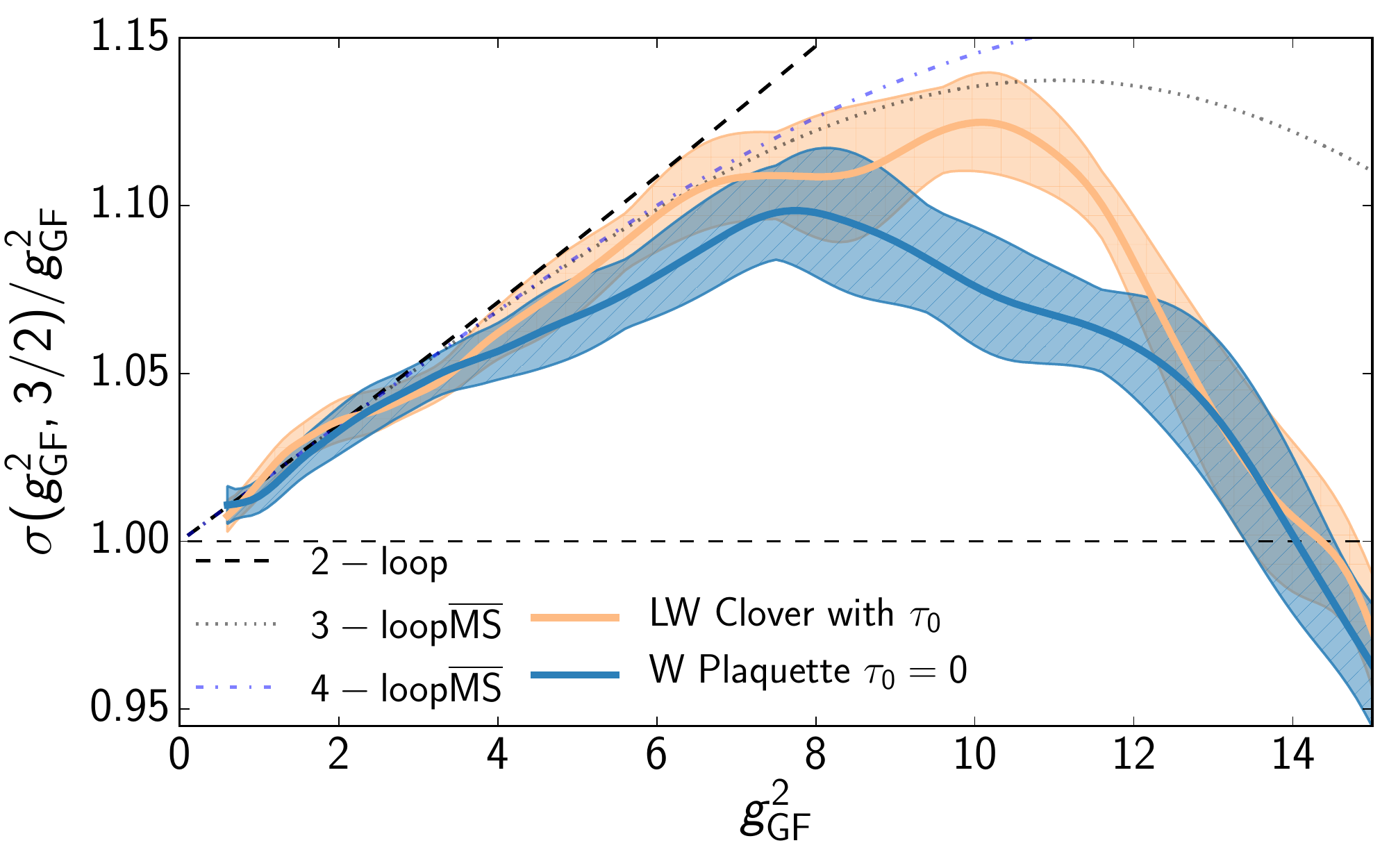}
  \includegraphics[width=0.49\textwidth]{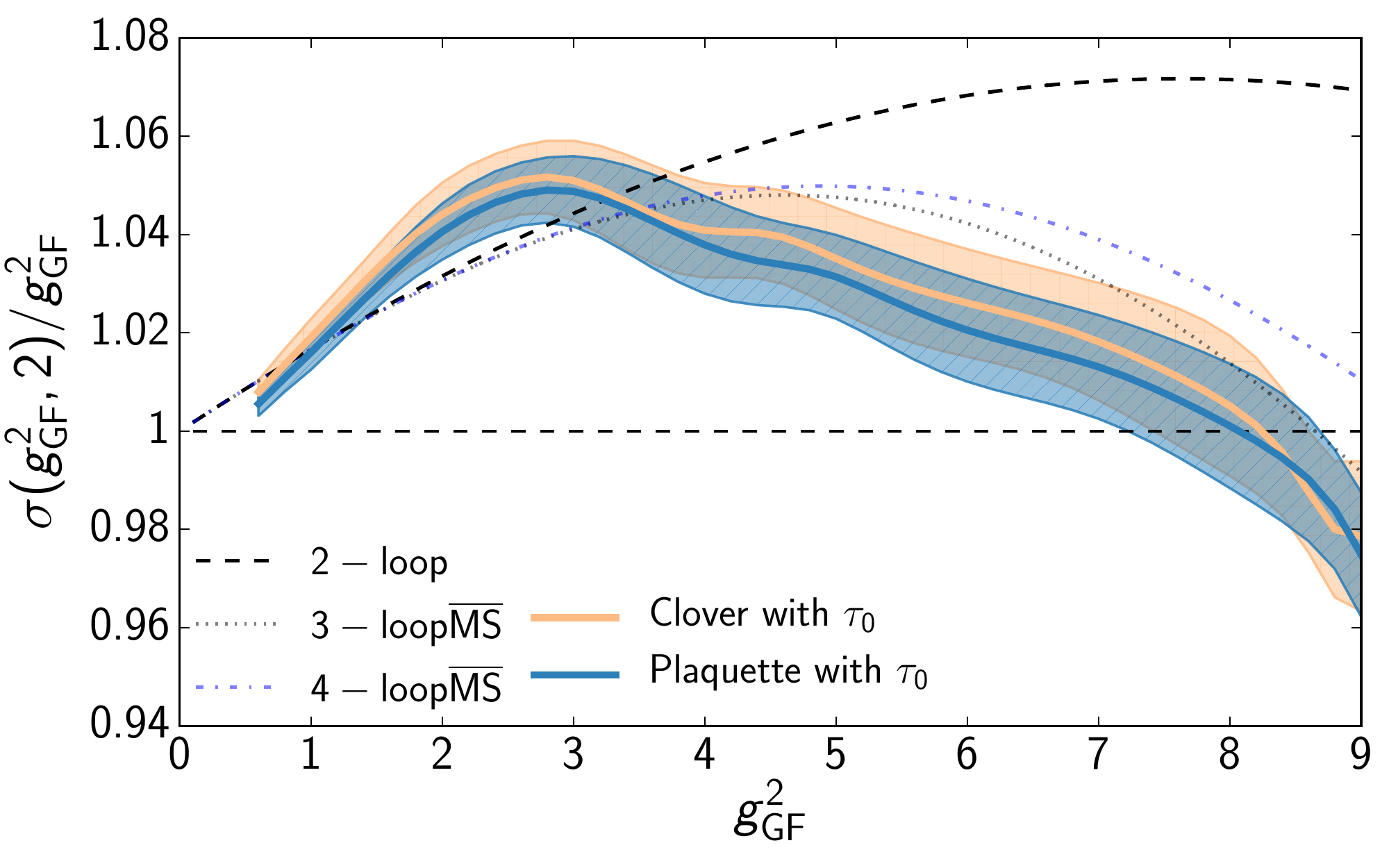} 
  \caption[b]{ 
  			  Continuum extrapolated step scaling function~\eqref{eq:stepscale} with different discretizations for:
              \emph{Left}: $N_f=6$ theory with $c_t=0.3$, $s=1.5$ and \emph{Right}: $N_f=8$ theory with $c_t=0.4$, $s=2$.
             }
  \label{kuva1} 
\end{figure}
In Fig~\ref{kuva1} we present the continuum extrapolated step scaling function~\eqref{eq:stepscale} for both the $N_f=6$ and $N_f=8$ fermion models.
In the $N_f=6$ model we use a step size of $s=3/2$ and set the scheme with $c_t=0.3$, while in the $N_f=8$ model we use a step size of $s=2$ and set the scheme with $c_t=0.4$.
In order to ensure correct continuum limit we perform the analysis using several different discretizations of flow and energy density, of which we show two examples with different colours in Fig~\ref{kuva1}.
We observe that our measurement follows the universal 2-loop $\MSb$ curve up to some intermediate coupling before diverting towards an IRFP. The higher order perturbative curves
are scheme dependent and only shown for a comparison. We do not plot the recent 5-loop result~\cite{Herzog:2017ohr} as not only does it not have an IRFP with these numbers of fermions,
it also develops two separate conformal windows in SU(2) models clearly indicating a breaking of perturbation theory at high couplings~\cite{Leino:2017hgm}.
As can be seen from Fig~\ref{kuva1}, the $N_f=6$ theory has an IRFP at $g_\ast^2=14.5(4)^{+0.4}_{-1.2}$~\cite{Leino:2017hgm} 
and the $N_f=8$ theory has an IRFP at $g_\ast^2=8.24(59)^{+0.97}_{-1.64}$~\cite{Leino:2017lpc}.
Here the first set of errors is the statistical error for the chosen set of discretizations and the second set gives the variation between different discretization choices.

\section{Anomalous dimension of the coupling}
\begin{figure}[t]
  \includegraphics[width=0.49\textwidth]{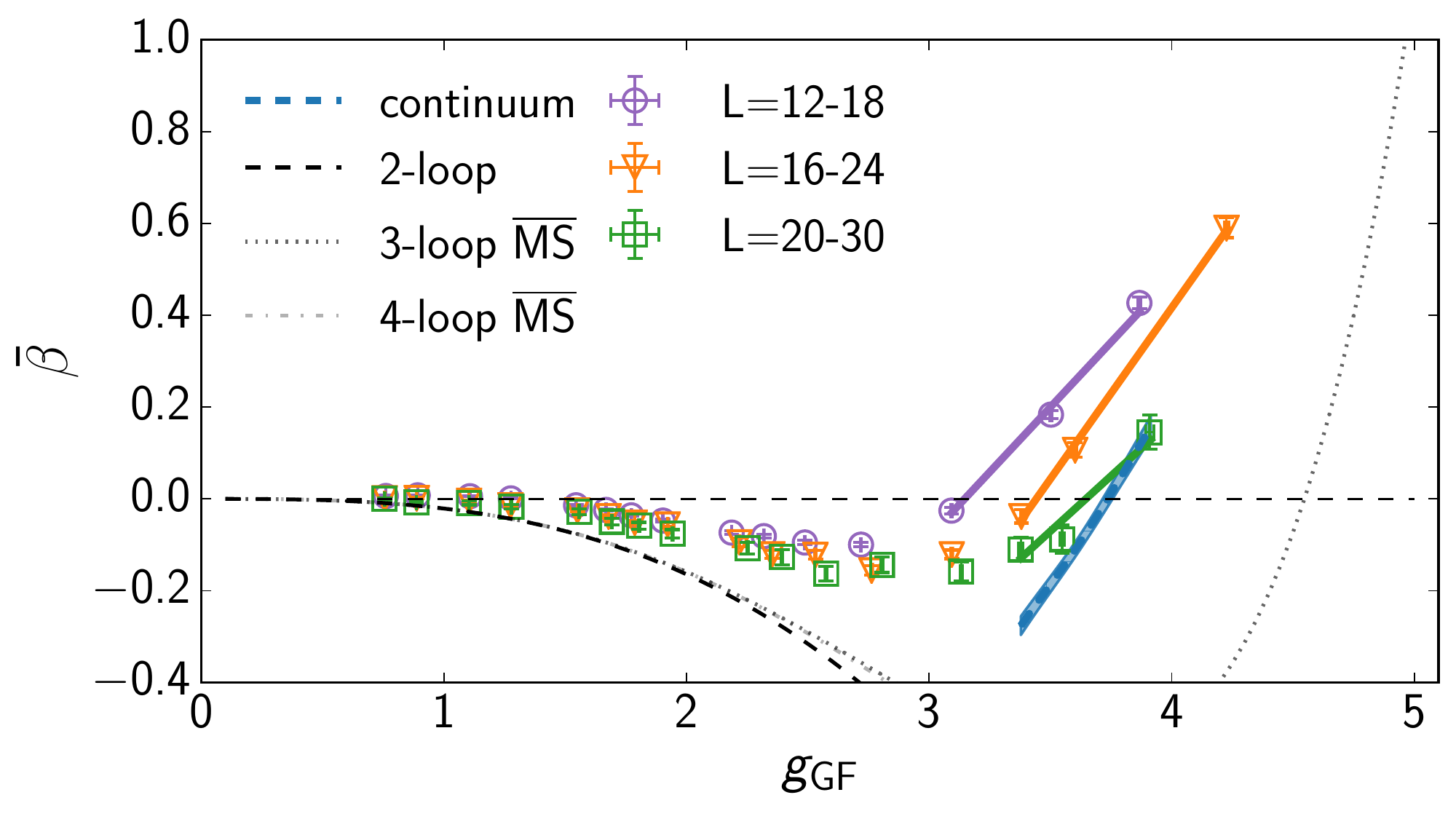}
  \includegraphics[width=0.49\textwidth]{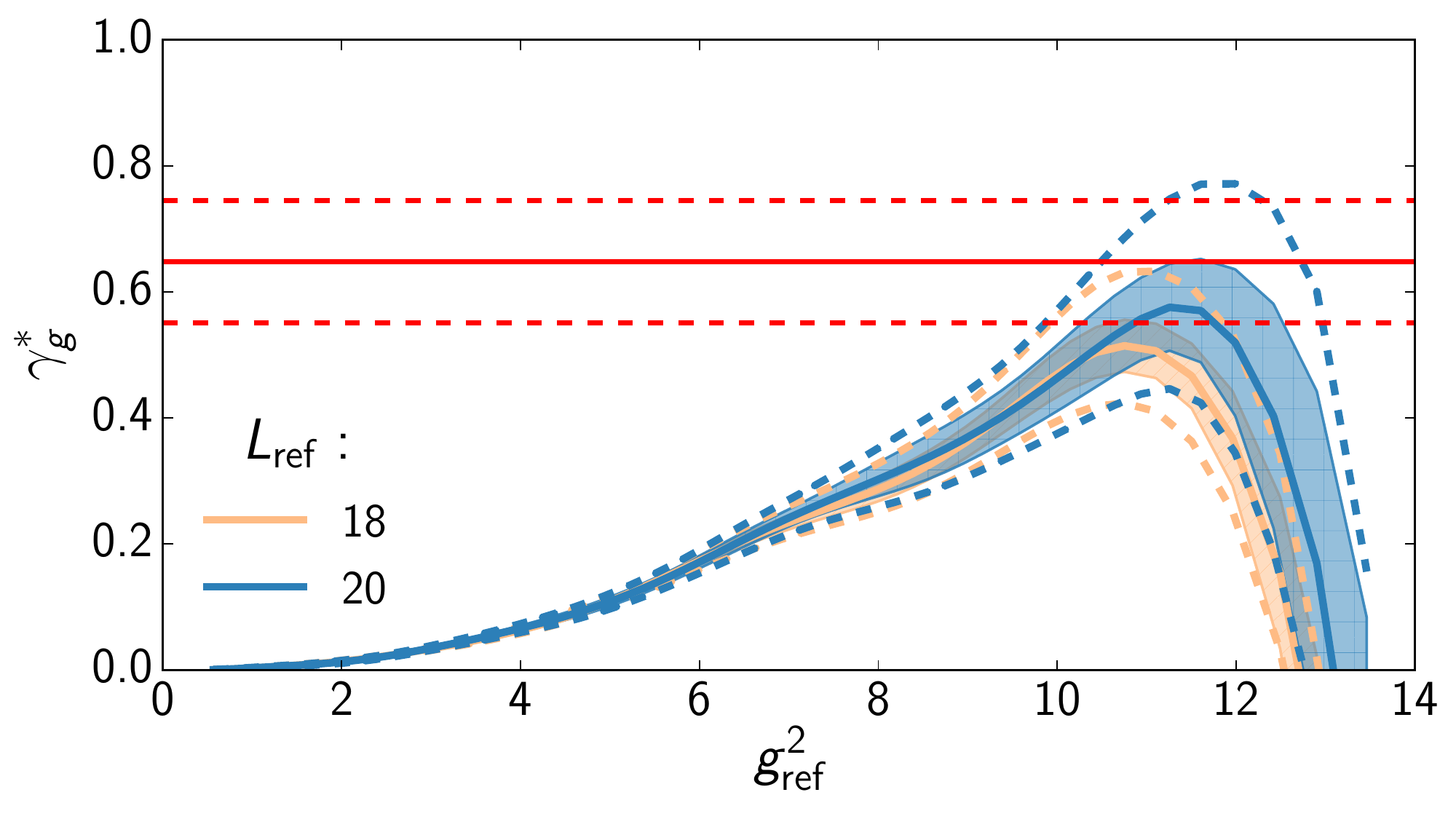} %ramos_norm.pdf} 
  \caption[b]{ 
              The anomalous dimension of the coupling $\gamma_g^\ast$ can be measured in two different ways: as a slope of the $\beta$-function \emph{(left)} 
			  or with finite-size scaling method \emph{(right)}. Both figures are for $N_f=6$ model. On the right hand figure, the red lines indicate the measurement from the slope,
			  and the difference between solid and dashed curves indicate the uncertainty in $g_\ast^2$.
             }
  \label{kuva2} 
\end{figure}
While the location of an IRFP is a scheme dependent quantity, the first irrelevant dimension of the gauge coupling $\gamma_g(g)=\beta^\prime(g)-\beta(g)/g$~\cite{NunesdaSilva:2016jfy} 
is invariant at the IRFP $g=g_\ast$. At the IRFP $\gamma_g(g_\ast)=\gamma_g^\ast$ reduces to a slope of the $\beta$-function, 
which we can measure directly from the step scaling function~\eqref{eq:stepscale}.
Alternatively, a finite-size scaling method of measuring $\gamma_g$ was proposed in~\cite{Appelquist:2009ty,DeGrand:2009mt,Lin:2015zpa,Hasenfratz:2016dou}.
In a functional form these two approaches can be written as:
\begin{equation}
\gamma_g^\ast=\frac{\mathrm{d}}{\mathrm{d}g}\frac{g}{2\mathrm{ln}(s)}\left.\!\left(1-\frac{\sigma(g^2,s)}{g^2}\right)\right\vert_{g=g_\ast}\,,\qquad
\gGF^2(\beta,L)-g_\ast^2 = \left[\gGF^2(\beta,L_\mathrm{ref})-g_\ast^2\right]\left(\frac{L_\mathrm{ref}}{L}\right)^{\gamma_g^\ast}\,.
\end{equation}
Both of these methods can be used with the same interpolated data used in section~\ref{sec:run}, but, because our interest now only lies at the IRFP, 
we can get statistically more significant result by doing lower order fit in the vicinity of the IRFP~\cite{Leino:2018qvq}. 
This procedure is demonstrated in the left side of Fig~\ref{kuva2} for the $N_f=6$ theory.

By measuring the slope of the $\beta$-function with multiple different fit ansatz we extract $\gamma_g^\ast=0.66(4)_{-0.13}^{+0.25}$ for the $N_f=6$ model and 
$\gamma_g^\ast=0.19(8)_{-0.09}^{+0.21}$ for the $N_f=8$ model~\cite{Leino:2018qvq}, where the two sets of errors are again the statistical and systematical errors, respectively.
These results are in agreement with the scheme independent perturbative estimates~\cite{Ryttov:2017kmx}: $\gamma_g^\ast=0.6515$ for $N_f=6$ and $\gamma_g^\ast=0.25$ for $N_f=8$.

On the other hand the finite-size scaling method offers a consistency check for the result obtained above. 
Because of the finite size effects and poor signal at the IRFP caused by $\gGF^2-g_\ast^2\sim0$, exact numbers cannot be extracted using this method.
On the right side of Fig~\ref{kuva2} we present this method for the $N_f=6$ model and we can see that the finite-size scaling method develops a maximum that agrees with the
measurement obtained from the slope of the $\beta$-function~\cite{Leino:2017hgm,Leino:2018qvq}.

\section{Mass spectrum}
\begin{figure}[t]
\centering
  \includegraphics[width=0.32\textwidth]{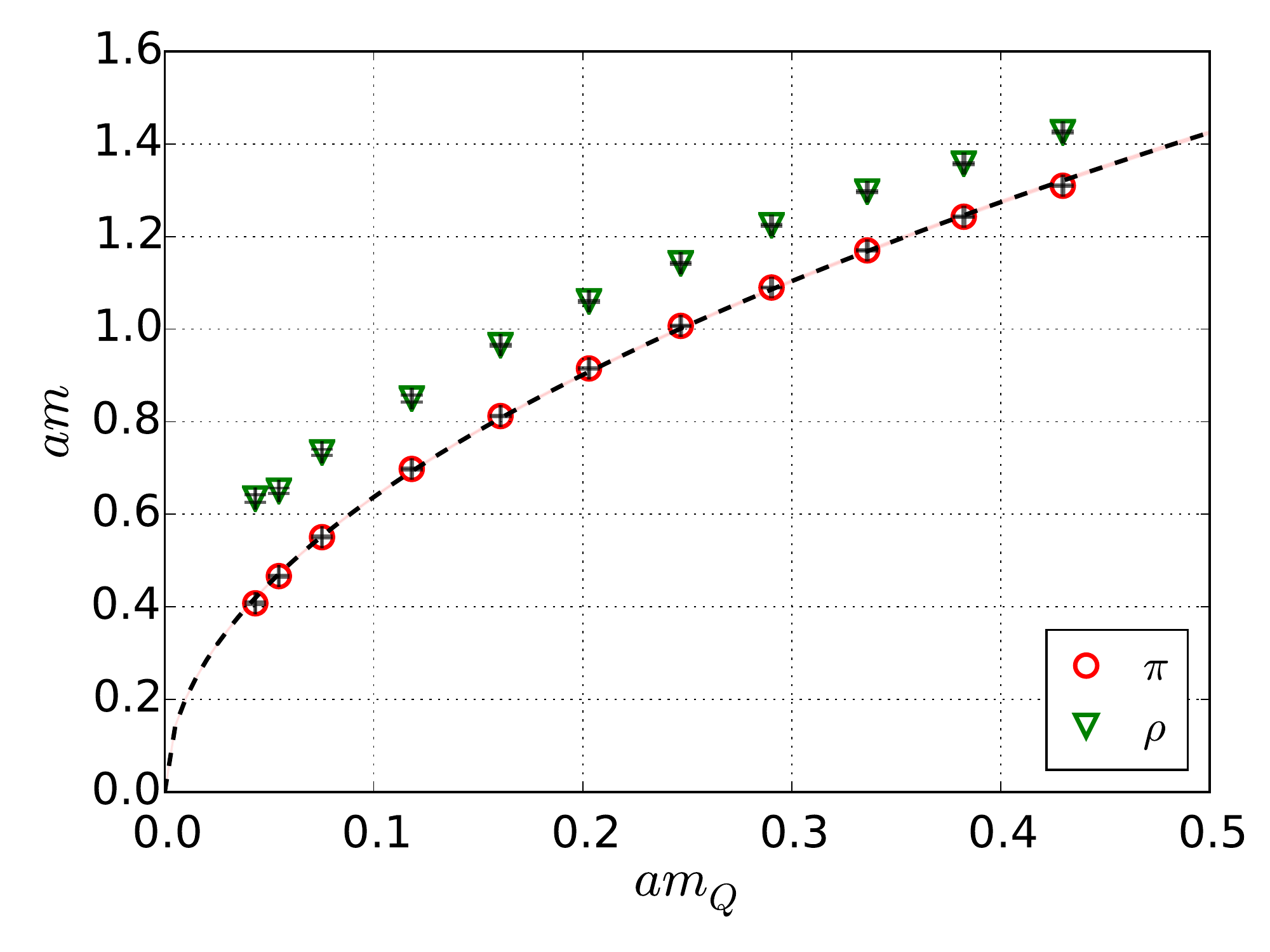}
  \includegraphics[width=0.32\textwidth]{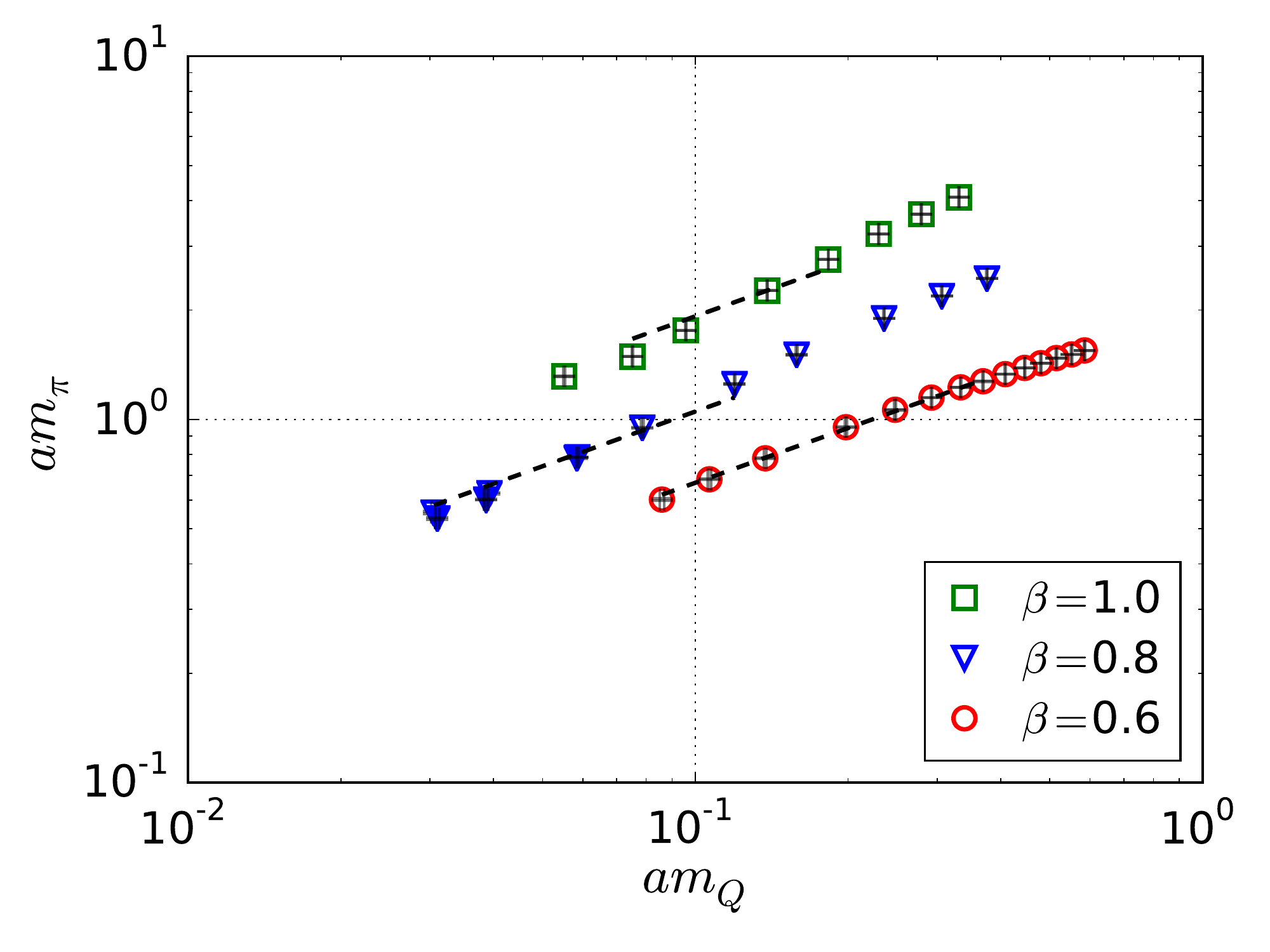} 
  \includegraphics[width=0.32\textwidth]{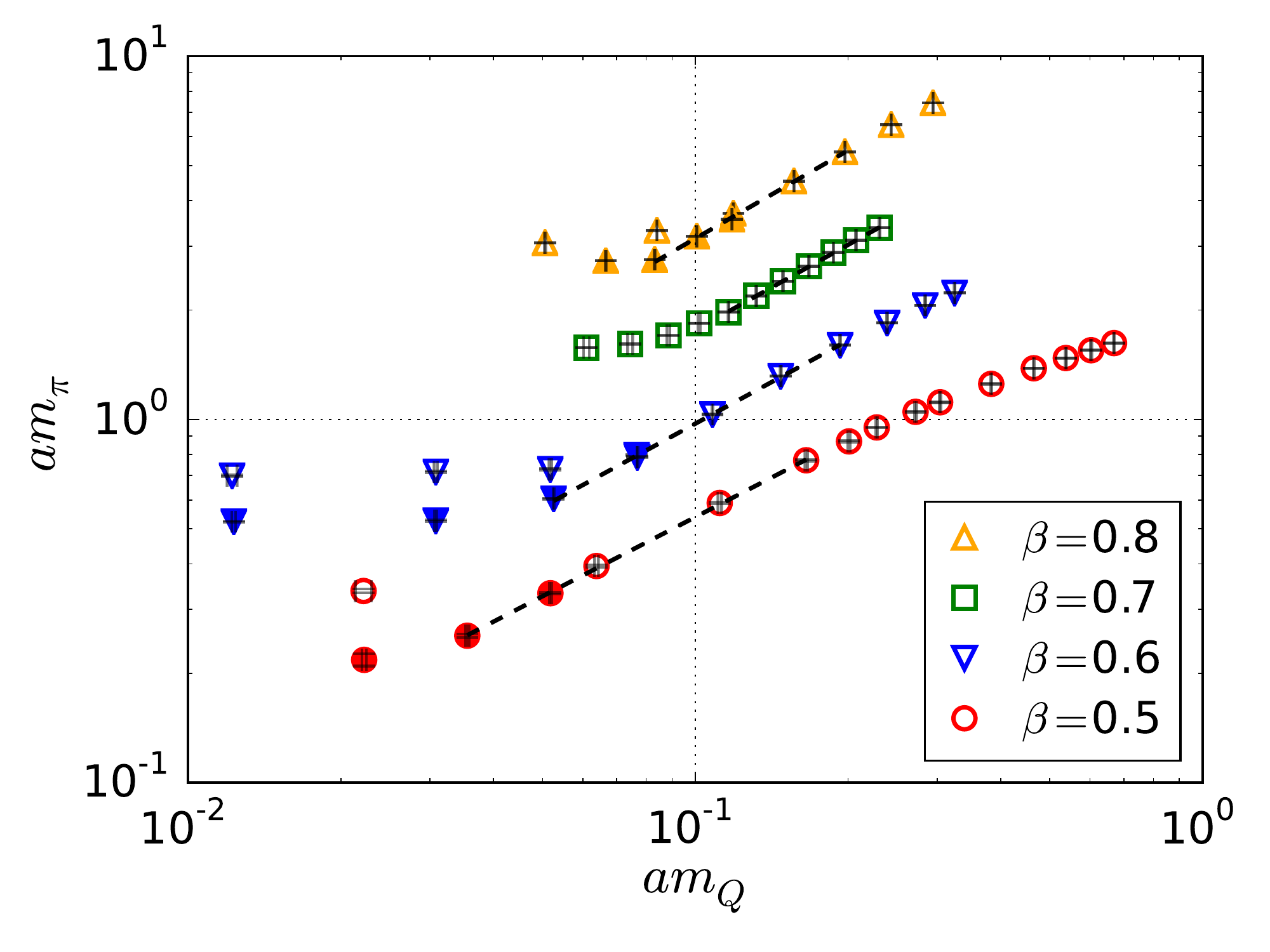}
  \includegraphics[width=0.32\textwidth]{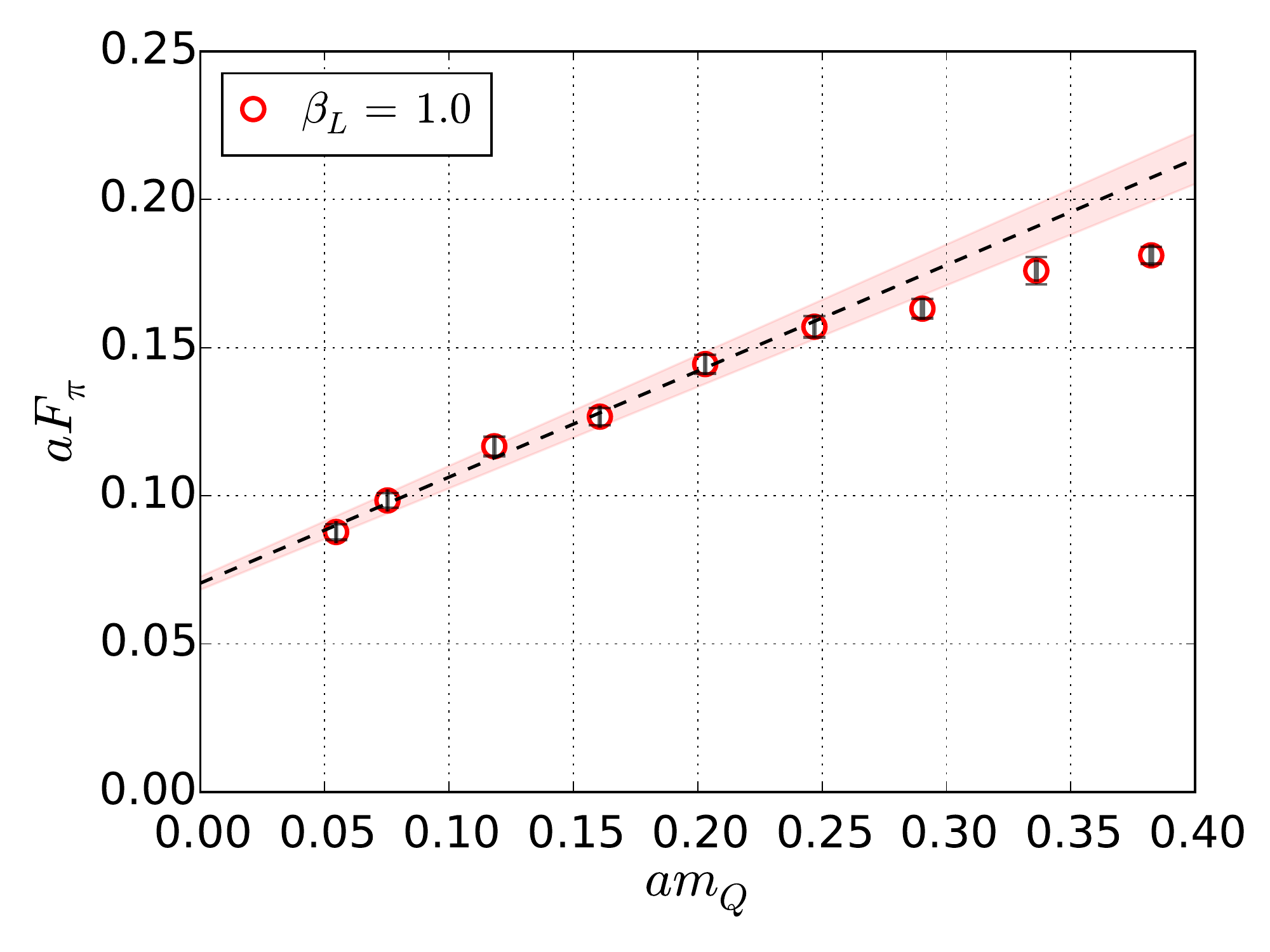}
  \includegraphics[width=0.32\textwidth]{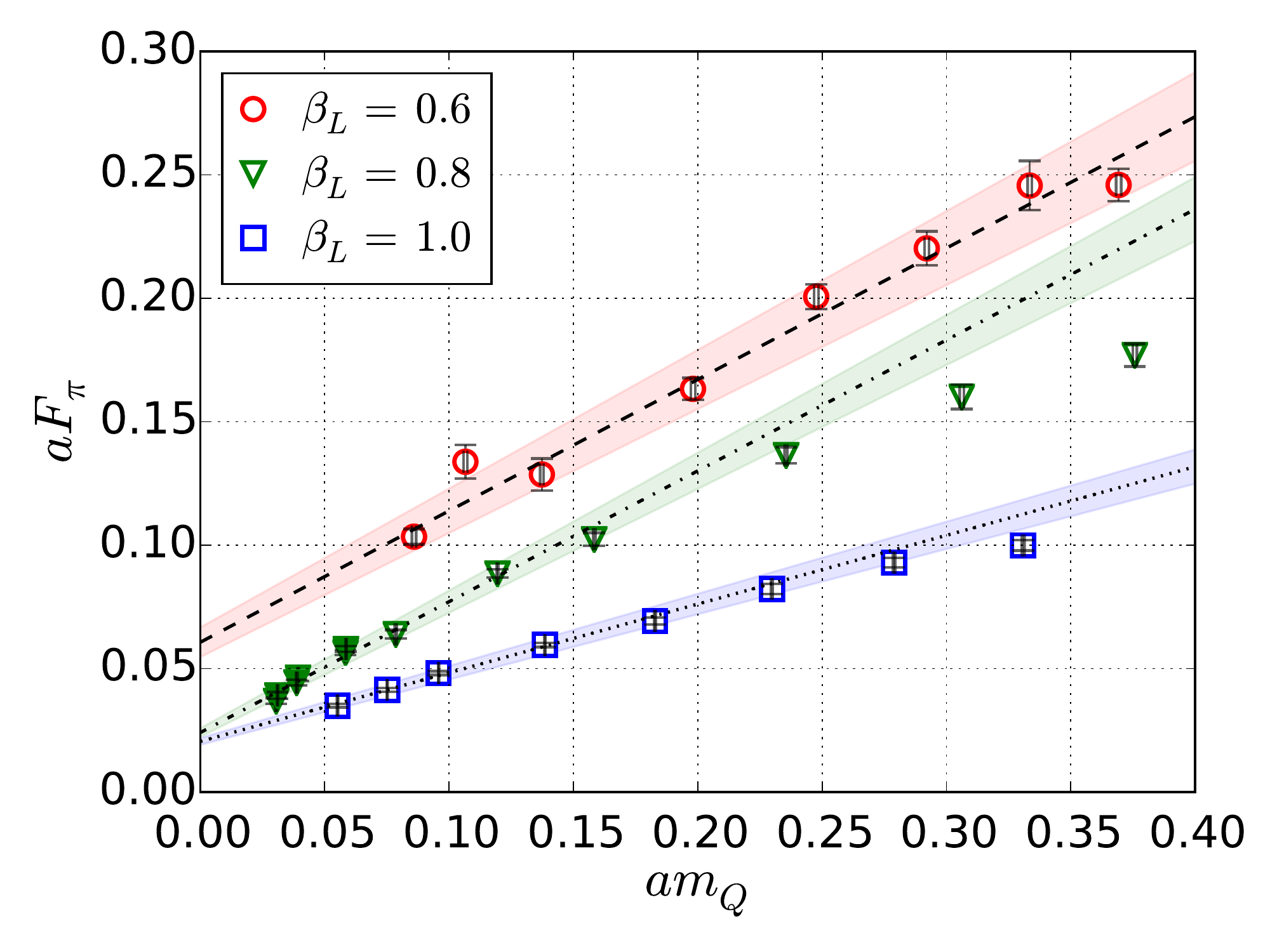}
  \includegraphics[width=0.32\textwidth]{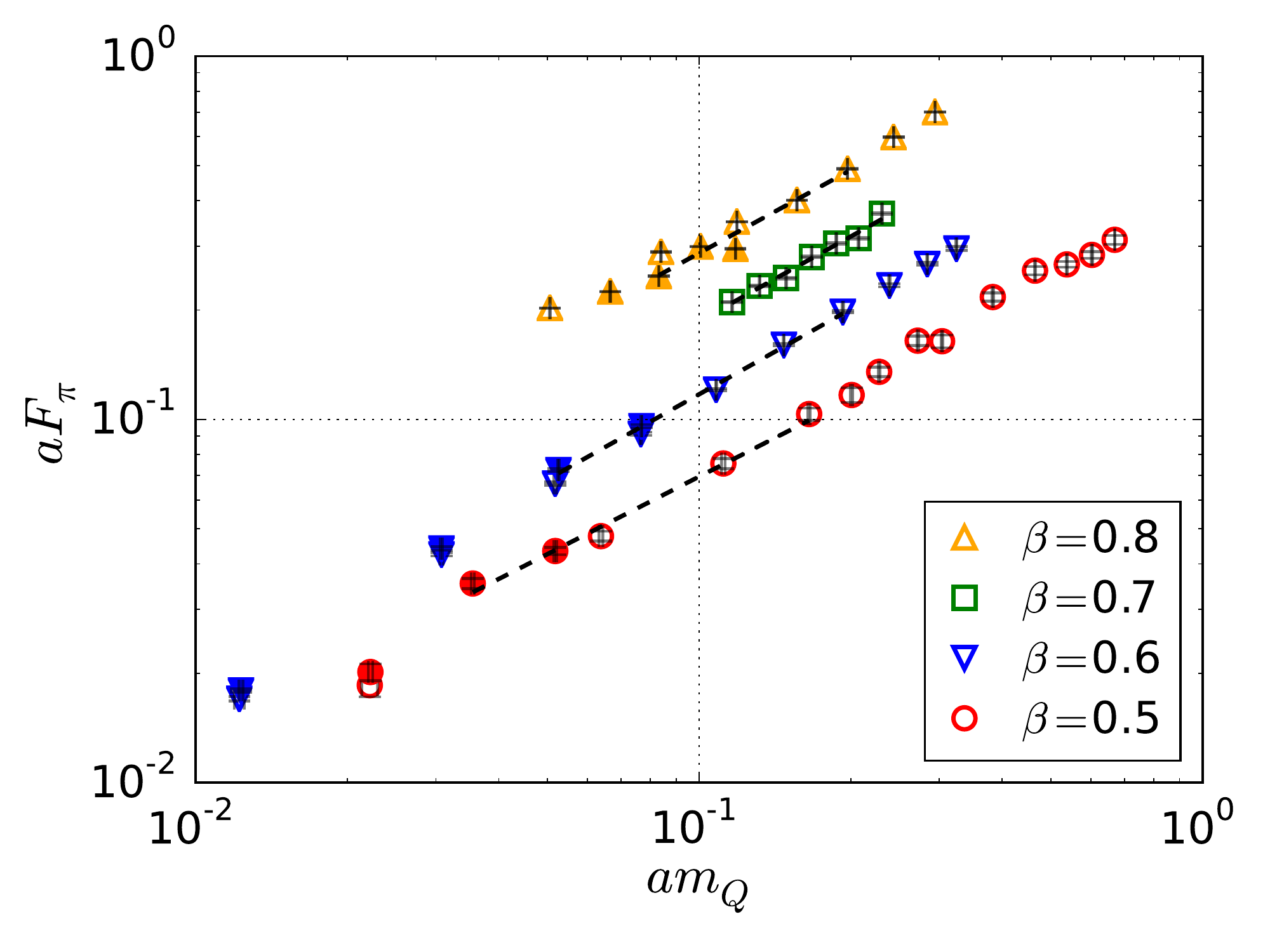} 
  \caption[b]{ 
             Singlet meson behaviour \emph{(upper row)} and pseudoscalar decay constants \emph{(lower row)} for 
			 $N_f=2$ \emph{(left)}, $N_f=4$ \emph{(middle)}, and $N_f=6$ \emph{(right)}.
             }
  \label{kuva4} 
\end{figure}
We have determined the masses of pseudoscalar $\pi$ and vector $\rho$ mesons by fitting time sliced average correlation functions with Coulomb gauge fixed wall sources.
The spectrum of these particles are shown in the upper row of Fig~\ref{kuva4} for the $N_f=2,4,6$ models. For the $N_f=4,6$ models we only show the $\pi$ meson, and for the $\rho$
results we refer the reader to the original paper~\cite{Amato:2018nvj}. As shown in the Fig~\ref{kuva4}, the $N_f=2$ model shows clear QCD-like chiral symmetry breaking behaviour
with the pseudoscalar scaling like $M_\pi\sim\sqrt{m_q}$ and the vector meson mass having a finite intercept in the $m_q\rightarrow0$ limit. The ratio $m_\rho/m_\pi$ diverges in this limit.

The $N_f=4$ model is closer to conformal window and therefore the signal is expected to be harder to measure. In Fig~\ref{kuva4} we observe the fit quality decreasing as the $\beta$ parameter
is increased, probably due to finite size effects. However, as long as the $\beta$ is reasonably small, we get a good square root fit for 
the $N_f=4$ model singlet meson behaviour which is observed to be similar to the $N_f=2$ case and consistent with a chiral symmetry breaking.
This confirms the previous running coupling result~\cite{Karavirta:2011zg}.

In the final $N_f=6$ case, we observe heavy volume dependence manifested by a plateau in meson measurements at small quark masses shown in Fig~\ref{kuva4}. 
As the difference between filled and hollow points show, the onset of the plateau moves to smaller quark masses as the volume is increased confirming this to be volume effect.
Regardless of these volume effects, it is still possible to do a scaling fit and extrapolate to zero mass. We observe that the pseudoscalar scales towards zero and
indicates a conformal dynamics, which confirms the result obtained from the running of the coupling~\cite{Leino:2017hgm}.

On the lower row of Fig~\ref{kuva4} we show the scaling behaviour of the pseudoscalar decay constant $F_\pi$ for all the models $N_f=2,4,6$. For the chirally broken cases $N_f=2$ and
$N_f=4$ we observe finite intercept of $F_\pi$ at zero quark mass limit while in the conformal $N_f=6$ model the $F_\pi$ scales to zero. For the numerical values of the interception points
we again refer the reader to the original paper~\cite{Amato:2018nvj}.
The pseudoscalar decay constant also offers a criteria to assess the finite volume effects. 
For the $N_f=2$ we have $F_\pi L > 1$ indicating the finite volume effects to be under control in the chiral limit. For the $N_f=4$ model we satisfy this condition only for the $\beta=0.6$ and
for the larger $\beta$ this onset of finite volume effects explains the decrease in fit quality in the upper row in Fig~\ref{kuva4}. For the $N_f=6$ model $F_\pi$ scales towards zero
indicating an IRFP.

\section{Mass anomalous dimension}
\begin{figure}[t]
\center
  \includegraphics[width=0.49\textwidth]{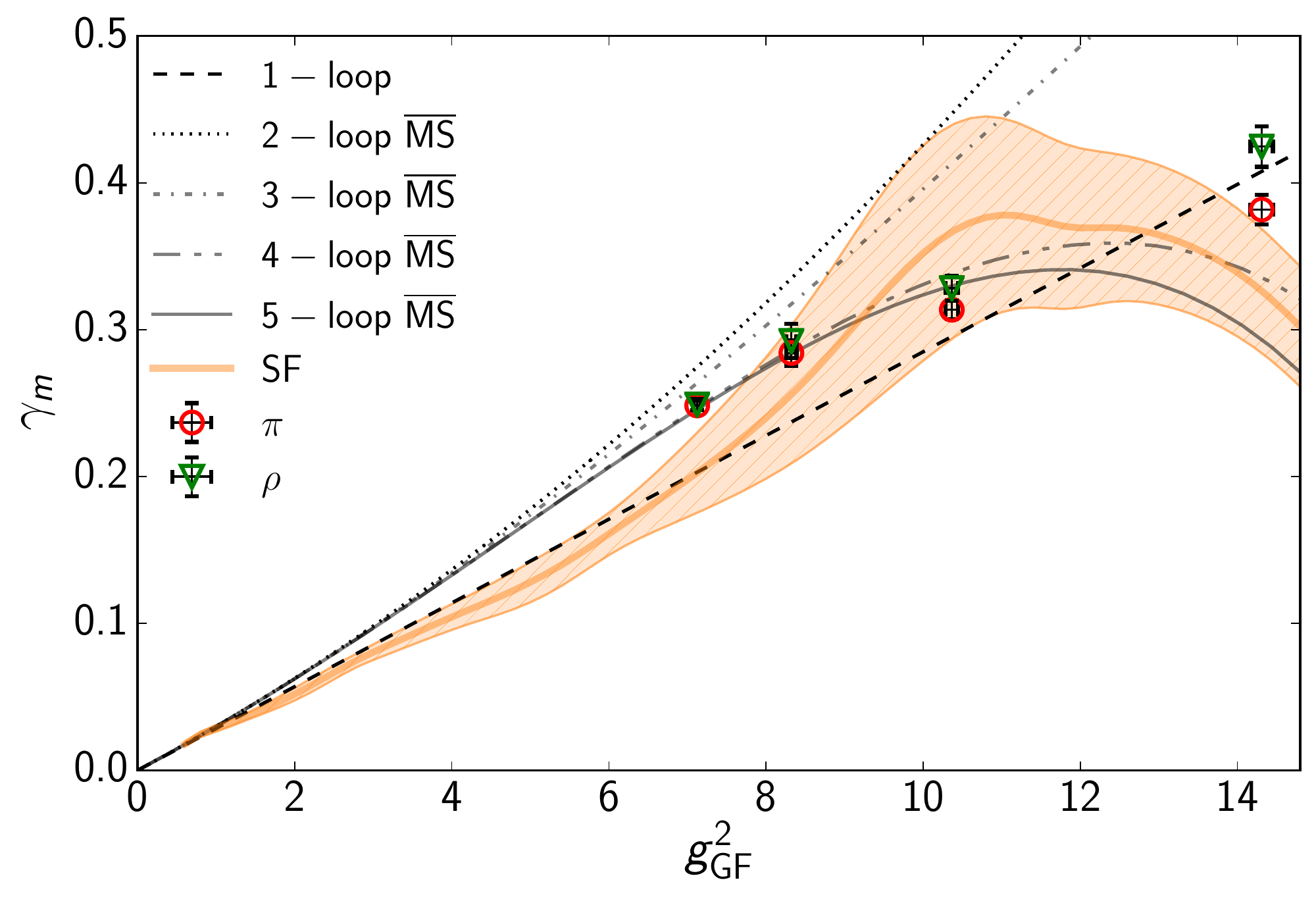}
  \includegraphics[width=0.36\textwidth]{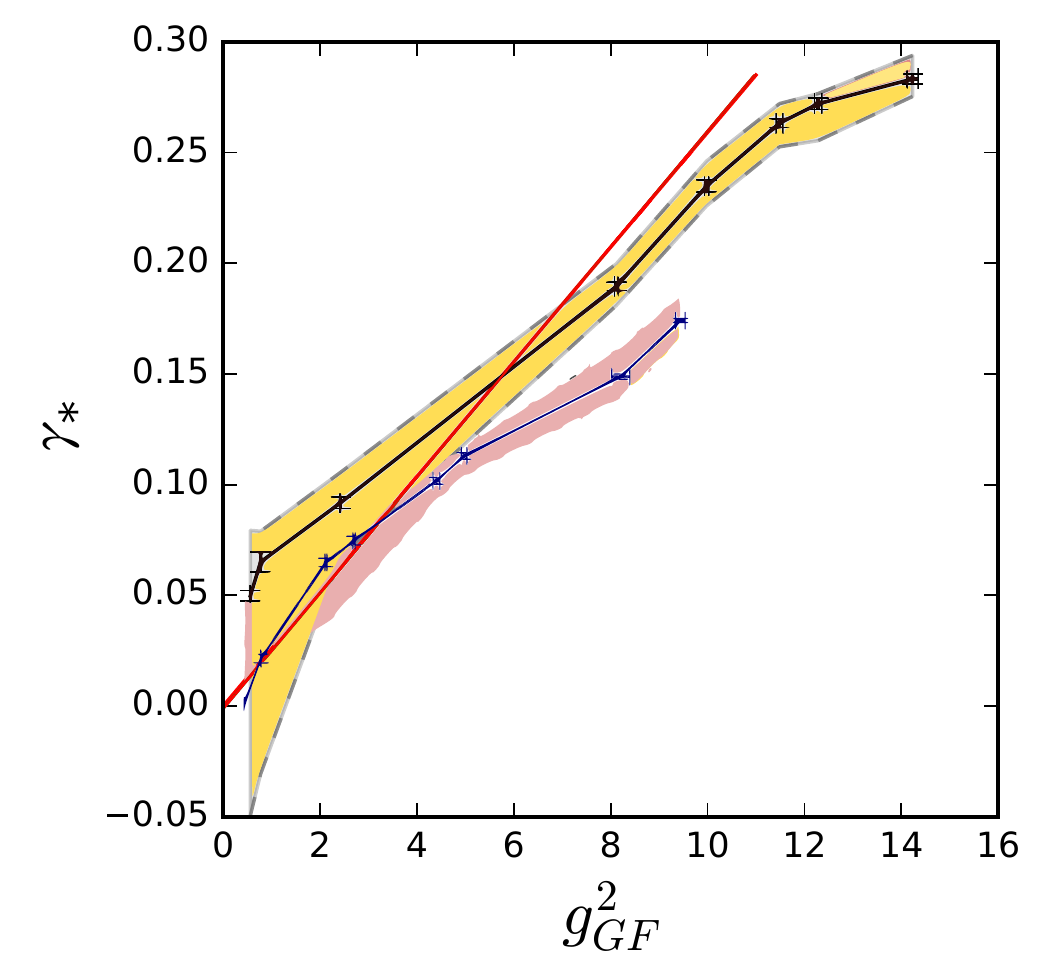} 
  \caption[b]{ 
              The mass anomalous dimension $\gamma_m$ with step scaling method (\emph{left, shaded curve}), scaling of meson masses (\emph{left, points}), and
			  spectral density method (\emph{right}). On the left we only have $N_f=6$ and on the right we have both $N_f=6$ in yellow-black and $N_f=8$ in pink-blue.
             }
  \label{kuva3} 
\end{figure}
As a gauge theory approaches an IRFP the coupling becomes and irrelevant direction and only relevant operator is the quark mass. 
At the IRFP, where $g^2=g^2_\ast$, all composite-state masses run to zero together with the quark mass
with an universal exponent: $m_q^{1/(1+\gamma_m^\ast(g_\ast))}$. We can fit this behaviour to both pseudoscalar masses in Fig~\ref{kuva4} and vector masses~\cite{Amato:2018nvj} to measure the mass anomalous
dimension $\gamma_m$. This measurement is shown as coloured points on the left side of Fig~\ref{kuva3}. 

Alternatively, we can measure the mass anomalous dimension from the configurations used in the running of the coupling study~\cite{Leino:2017hgm,Leino:2017lpc}. 
We extract the $\gamma_m$ using two different methods: the step scaling method and the spectral density method. In the step scaling method we measure the running of 
the pseudoscalar density renormalization constant
$Z_P$ and define $\gamma_m(u)\mathrm{log}(s)=-\lim_{a\rightarrow 0}\mathrm{log}[Z_p(\beta,sL/a)/Z_p(\beta,L/a)]|_{\gGF^2(\beta,L/a)=u}$,
where the continuum limit is taken assuming $\mathcal{O}(a^2)$ scaling. Interpolating $Z_p$ with 8th degree polynomial we get the continuum limit shown in Fig~\ref{kuva3} for the $N_f=6$ theory.
This method becomes rather unstable at higher couplings which is perceived as increased errors. Again the higher order perturbative curves are shown as a reference and are not directly comparable.
For $N_f=8$ theory the step scaling method is unstable and is not presented here; our take on the continuum limit in $N_f=8$ model can be found in~\cite{Leino:2017hgm}.

The last method of measuring $\gamma_m$ is the spectral method. In the spectral method we stochastically extract the mode number of Dirac operator~\cite{Giusti:2008vb,Patella:2011jr}.
The mode number is known to a scaling $\Lambda^{4/(1+\gamma_m^\ast}$ in the vicinity of an IRFP. With this method we can extract $\gamma_m$ for both $N_f=6$ and $N_f=8$ as shown on the
right side of Fig~\ref{kuva3}. We also observe that the results agree with those attained with other two methods. The only caveat of this method is that we only get a good signal at large lattices,
and hence we have not enough points for proper continuum limit and the curves shown in Fig~\ref{kuva3} are only for the single largest lattice sizes. 
We measure the values $\gamma_m^\ast=0.283(2)_{-0.01}^{+0.01}$ for $N_f=6$ and $\gamma_m^\ast=0.15(2)$ for $N_f=8$.

\section{Conclusions}\label{concl}
\begin{table}
\centering
\begin{tabular}{cccc|cc}
        & $\gGF^2$ & $\gamma_g^\ast$ & $\gamma_m^\ast$ & $\gamma_g^{\mathrm{R\&S}}$ & $\gamma_m^{\mathrm{R\&S}}$ \\
\hline
$N_f=6$ & $14.5(4)^{+0.4}_{-1.2}$  & $0.66(4)_{-0.13}^{+0.25}$ & $0.283(2)_{-0.01}^{+0.01}$ & $0.6515$ & 0.6 \\
$N_f=8$ & $8.24(59)^{+0.97}_{-1.64}$ & $0.19(8)_{-0.09}^{+0.21}$ & $0.15(2)_{-0.01}^{+0.01}$                  & $0.25$   & 0.3
\end{tabular}
\caption{Location of IRFP and its anomalous dimensions compared to scheme invariant perturbative estimates~\cite{Ryttov:2017kmx}.}
\label{tabtab}
\end{table}%
We have studied the running coupling and mass scaling in the SU(2) lattice gauge theory with 2--8 massless Dirac fermions. From the results it is clear that the $N_f=2$ and $N_f=4$ models are chirally broken
and have QCD-like behaviour. Meanwhile, we have presented strong evidence on the existence of an IRFP in $N_f=6$ and $N_f=8$ models. For these conformal models we have completed table~\ref{tabtab} 
with all the properties we have measured at the IRFP. We have also included comparison to the scheme independent perturbative estimation~\cite{Ryttov:2017kmx}.
We have checked the scheme independence of these results by varying the gradient flow parameter $c_t$, 
these numbers and many more details are available in the original publications~\cite{Leino:2017lpc,Leino:2017hgm,Leino:2018qvq,Amato:2018nvj}.

\section{Acknowledgements}
This work is supported by the Academy of Finland grants 310130, 308791 and 267286.  
S.T. is funded by the Magnus Ehrnrooth foundation. J.M.S. and V.L. have been funded by the Jenny and Antti Wihuri foundation.
V.L is currently supported by the DFG cluster of excellence “Origin and Structure of the Universe” and
Bundesministerium f\"ur Bildung und Forschung (BMBF) under the grant 
Verbundprojekt 05P2018 - Ausbau von ALICE am LHC: Hydrodynamische Entwicklung und Berechnung des nuklearen Modifikationsfaktors, Projekt Nr. 05P18WOCA1.
The simulations were performed at the Finnish IT Center for Science (CSC) in Espoo, Finland.

\bibliographystyle{jhep}
\bibliography{su2_nf6.bib}

\end{document}